\documentclass[10pt]{iopart}
\usepackage{lipsum,multicol}
\usepackage{graphicx}
\usepackage[utf8]{inputenc}
\usepackage[T1]{fontenc}
\usepackage{indentfirst}
\usepackage{tabularx,ragged2e,booktabs,caption}
\usepackage[font=footnotesize,labelfont=bf]{caption}
\usepackage{geometry}
\usepackage{color, colortbl}
\usepackage{multirow}
\usepackage{url}
\usepackage{amssymb}
\usepackage{ulem}
\usepackage{hyperref}
\makeatletter
\def\@bibitem#1{\item[]%
    \if@filesw\immediate\write\@auxout{\string \bibcite {#1}{\the\value{\@listctr }}}\fi\ignorespaces}
\makeatletter

\newenvironment{myindentpar}[1]%
{\begin{list}{}%
  {\setlength{\leftmargin}{#1}}%
  \item[]%
}
{\end{list}}

\definecolor{DGray}{gray}{0.8}
\definecolor{LGray}{gray}{0.9}

\begin{document}

\title{Characterization of the SPARC4 CCDs}

\author{D V Bernardes$^1$; E Martioli$^1$; C V Rodrigues$^2$.}

\address{$^1$Laboratório Nacional de Astrofísica, Rua Estados Unidos, 154, Itajubá-MG, Brasil}
\address{$^2$Instituto Nacional de Pesquisas Espaciais, Avenida dos Astronautas, 1758, São José dos Campos-SP, Brasil.}

{\footnotesize \hspace{50pt} E-mail: dbernardes@lna.br, emartioli@lna.br, claudia.rodrigues@inpe.br.}

\begin{abstract}
We present the photometric characterization of the four iXon Ultra 888 CCD cameras of the SPARC4 instrument, which will be installed on the 1.6 m telescope of the {\it Pico dos Dias} Observatory in Brazil. We applied experimental methodologies for a systematic characterization of the read noise, electronic gain, dark current, and quantum efficiency of the CCDs. We have analyzed the statistical distribution of the read noise, and also its spatial gradient and temporal variability, where we obtained an average value of the read noise of 6.33 electrons. We applied the Janesick method to determine the electronic gain, where we obtained an average value of 3.35 e-/ADU. We have also obtained an average dark current of 0.00014 e-pix\textsuperscript{-1}s\textsuperscript{-1} for CCD internal temperature of -70 ºC. We have inspected the dependency of the dark current with temperature and the spatial distribution of the dark current, where we found a variable profile in the CCD 9917. We developed an experiment using a bench mounted monochromator to obtain the spectral dependency of the quantum efficiency in the spectral range between 350 nm and 1100 nm, where we measured the quantum efficiency for each camera. The camera 9915 presents the highest quantum efficiency of 95.8 \%. Our results are compared with those from the manufacturer. These experiments allow us to diagnose the performance of these CCD cameras, an important sub-system of the SPARC4 instrument. It also provides a systematic way for monitoring the aging of the CCDs.

\end{abstract}
\maketitle

\section{Introduction}
\begin{multicols}{2}

The Simultaneous Polarimeter and Rapid Camera in 4 bands (SPARC4) (Rodrigues et al. 2012) is a new instrument being developed by the Astrophysics Division of the Instituto Nacional de Pesquisas Espaciais (INPE) in collaboration with the Laboratório Nacional de Astrofísica (LNA), in Brazil. SPARC4 will be installed on the 1.6~m Perkin-Elmer telescope at the Pico dos Dias Observatory (OPD), Brazil. The SPARC4 is an imager and polarimeter that makes use of three dichroic mirrors to divide the input light into four beams and hence to operate simultaneously in the following four Sloan Digital Sky Survey (SDSS) photometric bands (Gunn et al. 1998): \textit{ g, r, i}, and \textit{z}. Each of these beams will be recorded by an independent CCD device. The devices responsible for the data acquisition are the iXon Ultra 888 Electron Multiplying CCDs, produced by Andor Technology. The choice of these devices was based on their frame transfer and electron multiplying capabilities. The frame transfer allows faster acquisitions, providing an acquisition rate greater than 10 Hz, which is a requirement of SPARC4; the electron multiplying provides higher sensitivity, which allows the observation of relatively faint targets at high acquisition rates. Moreover, these devices present significantly low levels of noise and almost negligible dark current (Andor Technology 2015). Each SPARC4 camera has coating and window chosen according to the working spectral band. In addition, \textit{i} and \textit{z} cameras have fringe suppression treatment.

These iXon Ultra CCDs devices are configurable, allowing a diversity of operational modes, each providing a specific performance (sensitivity, noise, acquisition rate, etc.). Thus, the users can choose the proper operational mode for their science cases.

This paper presents the experiments developed to measure the performance parameters for the four SPARC4 iXon Ultra 888 CCD cameras, controlled by the Software Andor SOLIS, version 4.27.30001.0. Throughout the paper the cameras are also identified by their serial numbers and/or SPARC4 spectral channels: 9914 (g-channel), 9915 (r-channel), 9916 (i-channel), and 9917 (z-channel). We used the smallest vertical shift speed and pre-amplification options. We have limited the scope of our analysis to the following four parameters: the read noise, dark current (DC), electronic gain, and quantum efficiency (QE). The methodology to characterize each of these parameters and the results obtained are presented in Sections \ref{sec:RNcharact}-\ref{sec:QECharact}. The conclusions are presented in Section \ref{sec:conclusao}.

\section{Read noise} \label{sec:RNcharact}

To characterize the electronic read noise of the CCDs, we inspected the following properties of the bias level: its statistical distribution, spatial distribution, and temporal variability. The read noise is determined from a series of bias images, i.e., images obtained with the minimum exposure time and with the shutter closed. The CCD configuration used in this experiment is the following:

\begin{myindentpar}{0.5cm}
1. Shutter constantly closed \newline
2. Temperature of -70 ºC \newline
3. Exposure time of 1.0 x 10$^{-5}$ s \newline
4. Pre-amplification of 1\newline
5. Readout rate of 1 MHz\newline
6. Vertical shift speed of 4.33 x 10$^{-6}$ s \newline
7. Conventional Mode \newline
8. Kinetic mode with a kinetic \newline
cycle time of 12.0008 s\newline
\end{myindentpar}

\begin{figure*}
\centering
\includegraphics[scale=0.55]{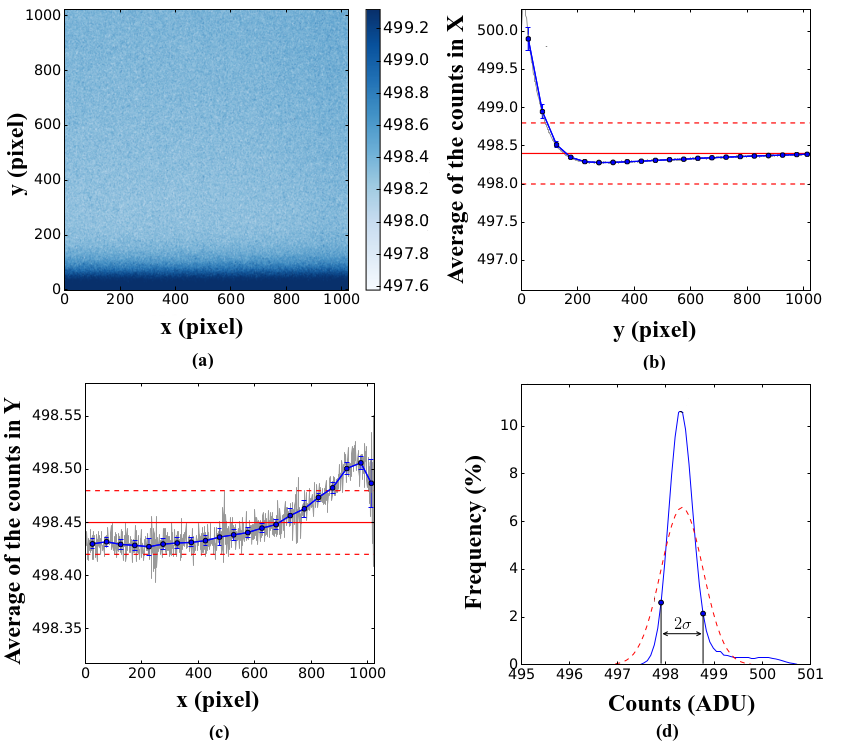}
\captionof{figure}{Panel (a) shows the average of 70 bias images obtained with camera 9914. Panels (b) and (c) show the mean of the rows along the \textit{y}-direction and the mean of the columns along the \textit{x}-direction, respectively. Gray lines show the mean values along the respective CCD direction (\textit{x} or \textit{y}). Blue lines show the binning of the mean series. Red lines present the mean (central line) and the one-sigma values (dashed lines). Panel (d) shows statistical distribution of the counts in panel (a). Blue solid line shows the probability distribution function, and dashed red curve shows the calculated normal probability distribution function.}
\label{fig:RN9914}
\end{figure*}

\begin{figure*}
\centering
\includegraphics[scale=0.5]{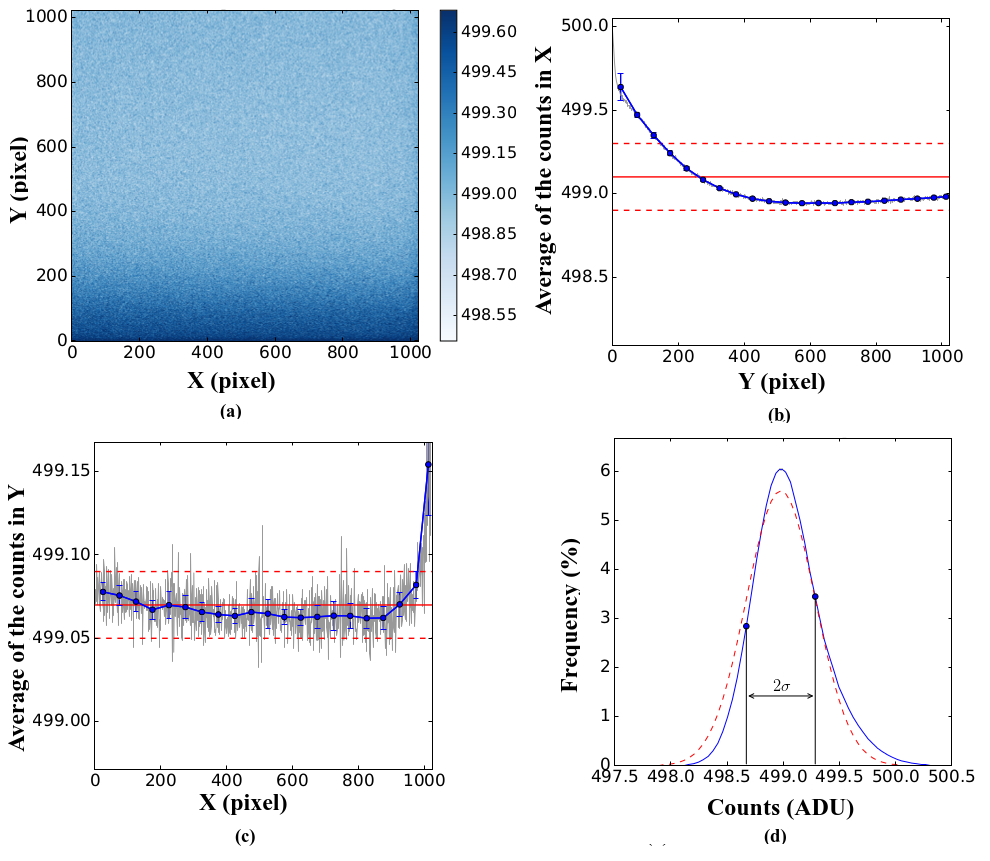}
\captionof{figure}{Same as Figure \ref{fig:RN9914} for camera 9915.}
\label{fig:RN9915}
\end{figure*}

\begin{figure*}
\centering
\includegraphics[scale=0.55]{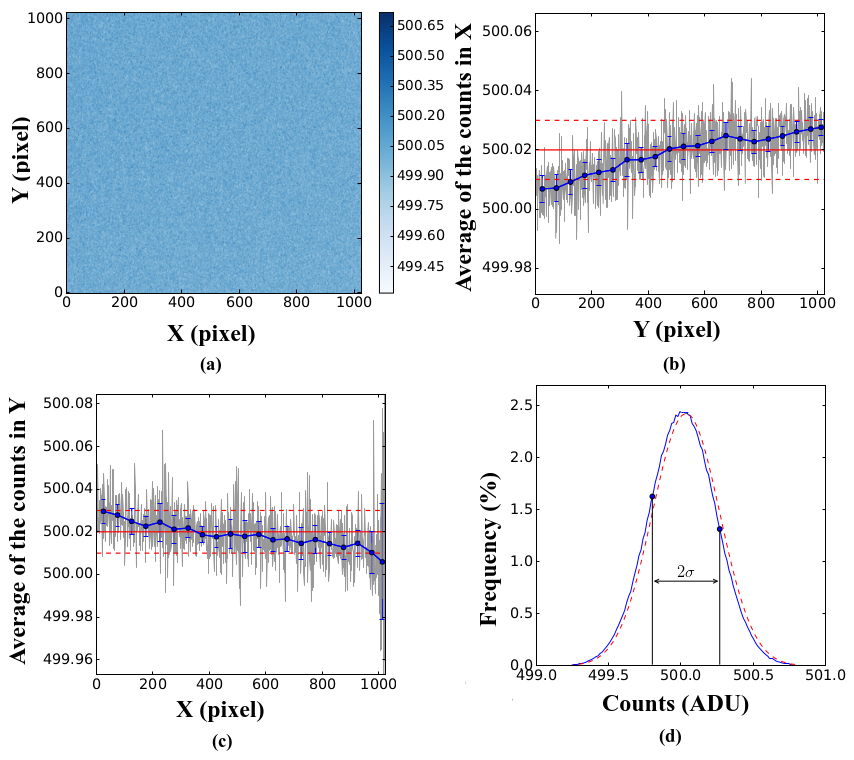}
\captionof{figure}{Same as Figure \ref{fig:RN9914} for camera 9916.}
\label{fig:RN9916}
\end{figure*}

\begin{figure*}
\centering
\includegraphics[scale=0.55]{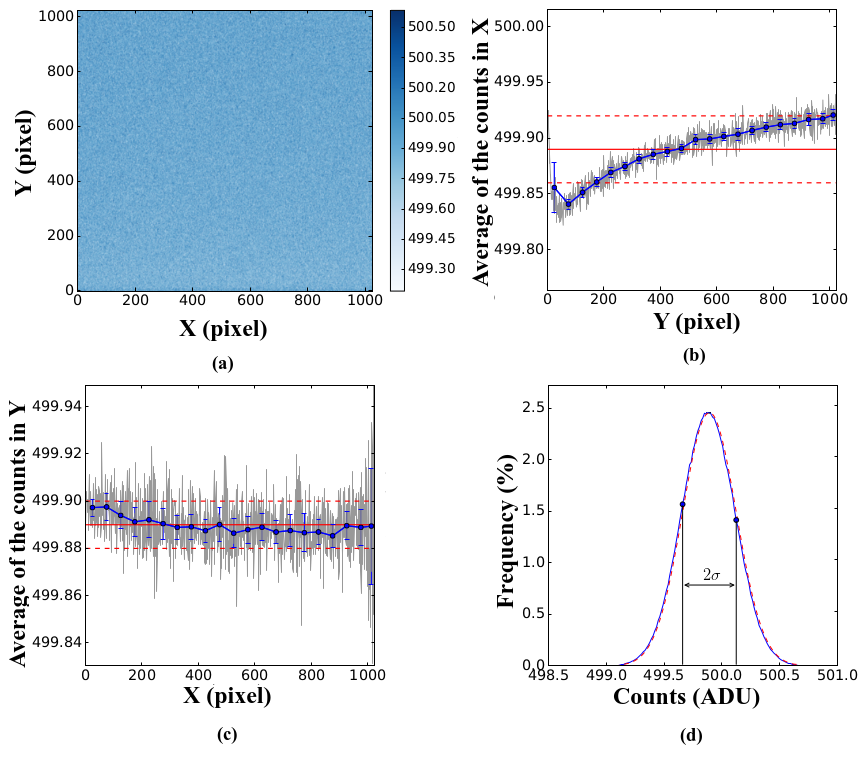}
\captionof{figure}{Same as Figure \ref{fig:RN9914} for camera 9917.}
\label{fig:RN9917}
\end{figure*}

Figures \ref{fig:RN9914}-\ref{fig:RN9917} present the results of the spatial distribution and the probability distribution function (PDF) of the bias obtained for the cameras 9914, 9915, 9916, and 9917, respectively. First, we combined 70 bias images by the mean, using a sigma clipping technique (Akhlaghi 2017) to eliminate outliers, as presented in panel (a). 

To inspect the spatial distribution of the bias level, we plot the mean of all CCD rows (collapse pixels in the \textit{x}-direction) as a function of CCD row number (\textit{y}-axis)\footnote{This is the CCD readout direction.}, and the mean of all CCD columns (collapse pixels in the \textit{y}-direction) as a function of CCD column number (\textit{x}-axis), as presented in panels (b) and (c), respectively. The spatial distribution of the bias presents a region of higher counts at the bottom of the CCDs 9914 and 9915. Each count represents a pixel value in analogical-to-digital unit (ADU).

We used the combined image of panel (a) to calculate the PDF of the counts, as presented in panel (d). We perform measurements of the mean ($\bar B_s$) and the standard deviation ($\sigma_s$) of this PDF (see Table \ref{tab:resultadoGradiente}), which corresponds to the mean bias level and the noise of a bias image, respectively. This noise can be composed by the read noise and a systematic component caused by spatial variations of the bias. To verify the normality of this PDF, we adjust a normal PDF using the average and standard deviation of the combined image. Notice that the PDF for cameras 9914 and 9915 deviates from the calculated PDF, which indicates the presence of high systematics.

\begin{figure*}
\centering
\hspace{0pt}\includegraphics[scale=0.4]{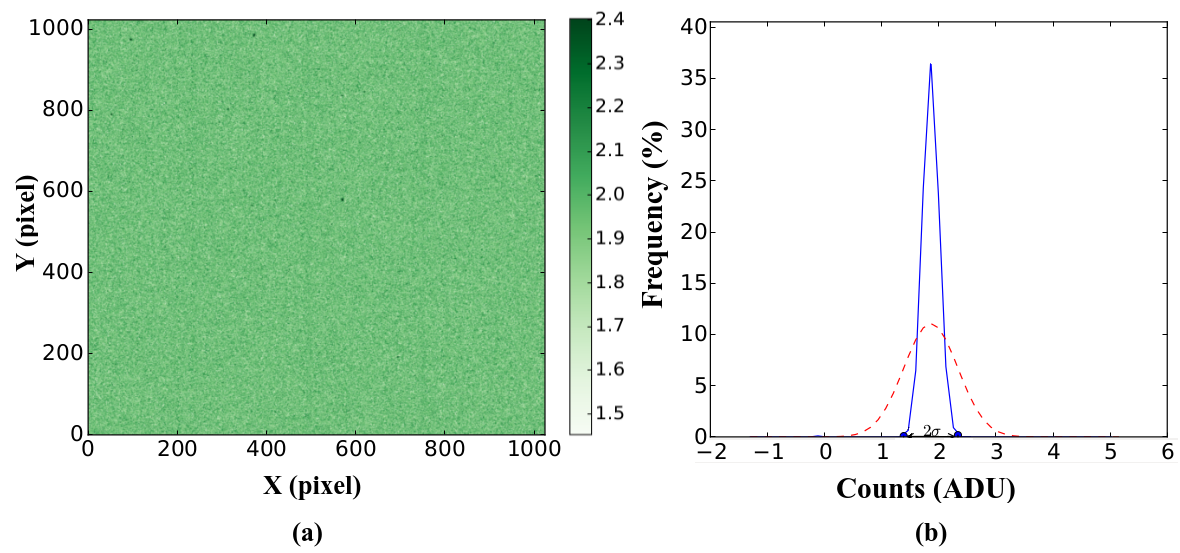}
\captionof{figure}{Panel (a) presents the pixel by pixel read noise distribution from a series of 100 bias images for CCD 9914. Panel (b) presents the probability distribution function of the read noise.  Blue solid line presents the probability distribution function of image (a), and red dashed lines present a normal probability distribution function calculated from the average and standard deviation of the image in panel (a).}
\label{fig:RNspatial9914}
\end{figure*}

\begin{figure*}
\centering
\hspace{0pt}\includegraphics[scale=0.35]{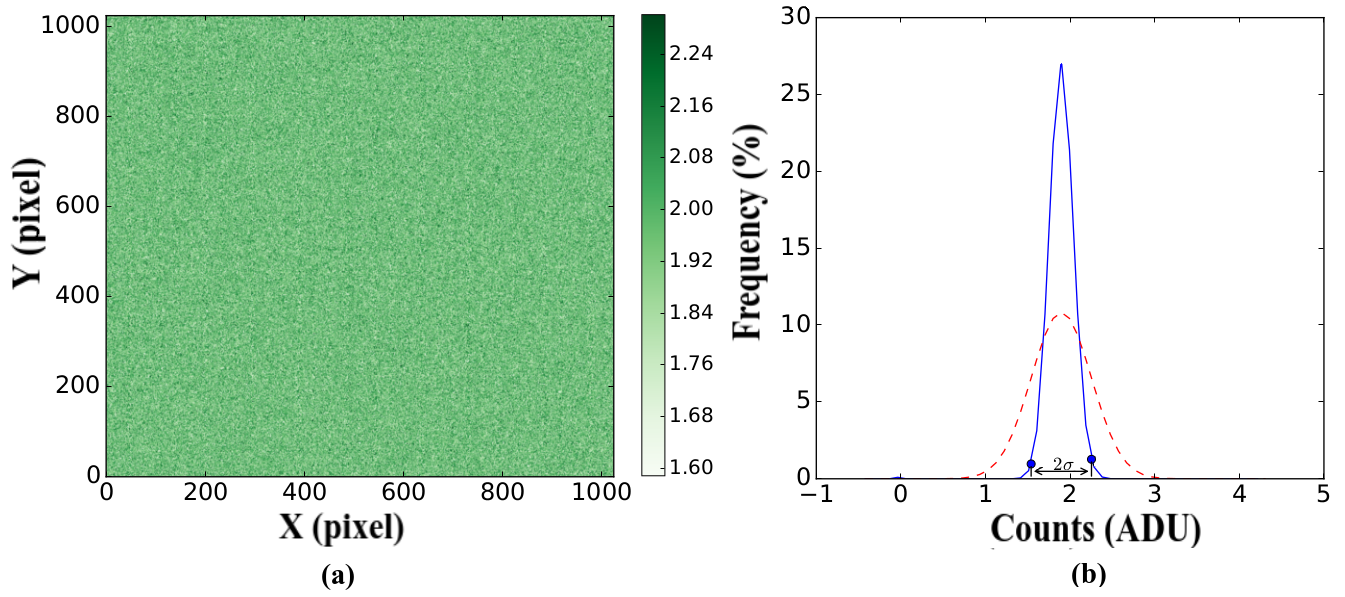}
\captionof{figure}{Same as Figure \ref{fig:RNspatial9914} for camera 9915.}
\label{fig:RNspatial9915}
\end{figure*}

\begin{figure*}
\centering
\hspace{0pt}\includegraphics[scale=0.4]{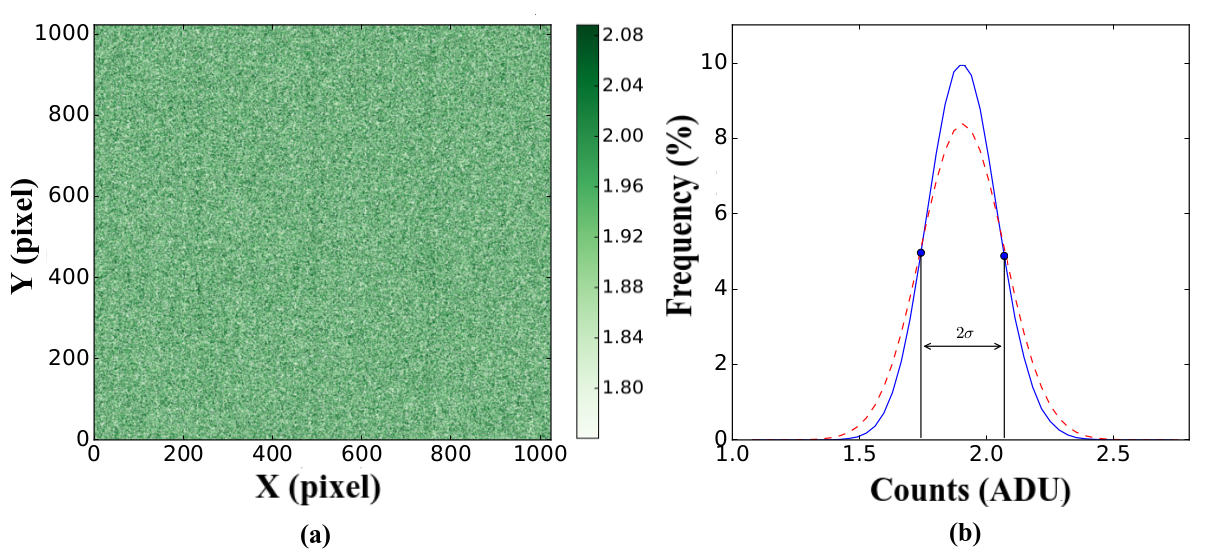}
\captionof{figure}{Same as Figure \ref{fig:RNspatial9914} for camera 9916.}
\label{fig:RNspatial9916}
\end{figure*}

\begin{figure*}
\centering
\hspace{0pt}\includegraphics[scale=0.4]{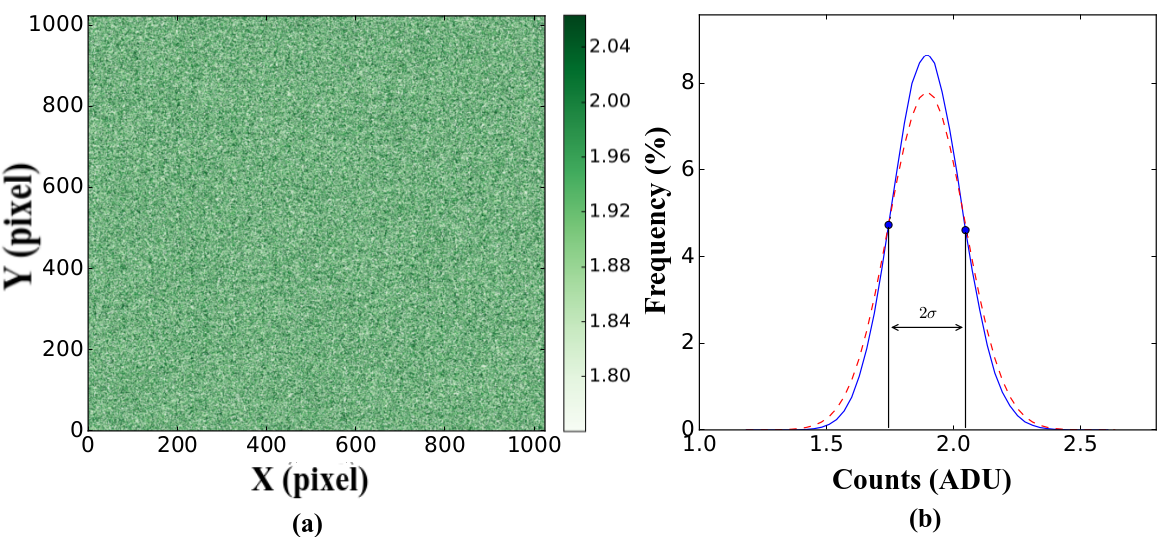}
\captionof{figure}{Same as Figure \ref{fig:RNspatial9914} for camera 9917.}
\label{fig:RNspatial9917}
\end{figure*}

Therefore, to obtain a more robust estimate of the read noise, we calculate the standard deviation of counts for each pixel in a set of 100 images, resulting in a noise image, as presented in panel (a) of Figures \ref{fig:RNspatial9914}-\ref{fig:RNspatial9917}. Panel (b) presents the calculated PDF of the read noise distribution of the panel (a) and a fit normal PDF. From the calculated PDF, we obtain the mean read noise value ($\bar M_{r}$) and its standard deviation (see Table \ref{tab:resultadoGradiente}). The read noise is apparently homogeneous along the entire CCD for all cameras, although, the PDF of the read noise deviates from the normal behavior, mainly for cameras 9914 and 9915, also indicating the presence of systematics.

\begin{figure*}
\hspace{-40pt}\includegraphics[scale=0.27]{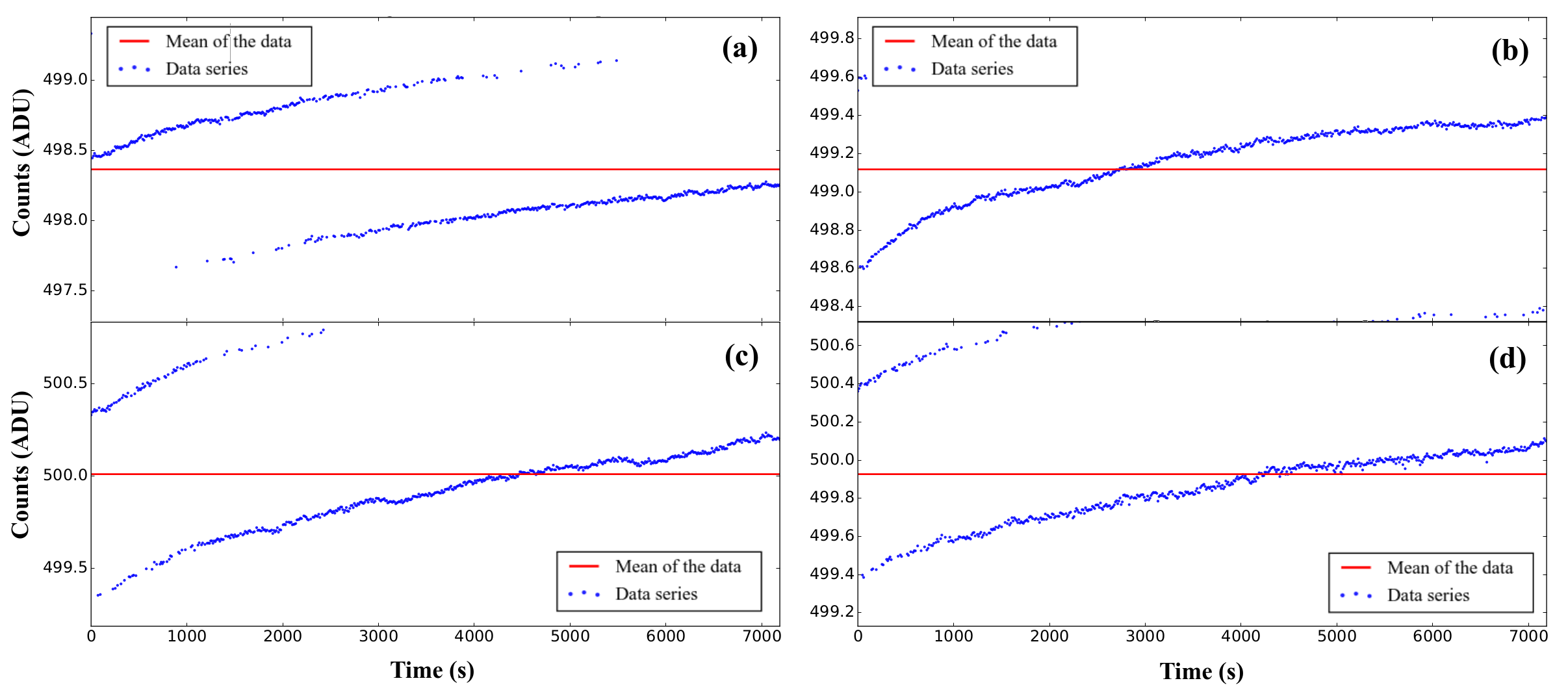}
\captionof{figure}{Average of each image for a series of 600 bias images as a function of the time for the cameras 9914 (a), 9915 (b), 9916 (c), and 9917 (d).}
\label{fig:resultadoBiasTemporal1}
\end{figure*}

To verify the temporal variability of the bias, we have acquired a series of 600 bias images, one every 12 s, for a time interval of two hours. Figure \ref{fig:resultadoBiasTemporal1} shows, for the set of cameras, the average of each image as a function of time, where the red line represents the average ($\bar B_t$) of this series (see Table \ref{tab:resultadoGradiente} for the values of $B_t$ and its dispersion). In this figure, it is possible to verify, for all CCDs, two regimes of counts with two different mean values. A possible explanation for this behavior is an electronic round of the counts for each pixel done by the camera in data digitalization. Both regimes presented an increase with time, which is probably due to the CCD heating, caused by energy dissipation of the charge transport. In our experiment, we found an average bias drift for all four CCDs of approximately 0.6 ADU in two hours.

\begin{figure*}
\centering
\hspace{-40pt}\includegraphics[scale=0.35]{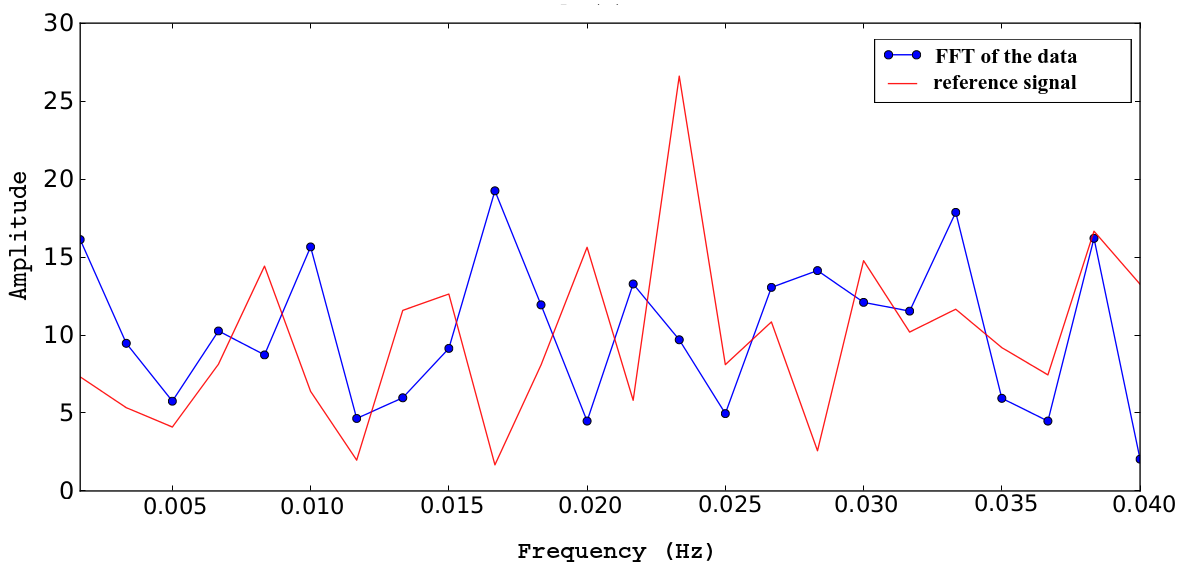}
\captionof{figure}{Fast Fourier Transform of the bias images series for camera 9914 (Figure \ref{fig:resultadoBiasTemporal1} (a)).}
\label{fig:resultadoBiasTemporal2}
\end{figure*}

We also calculated the Fast Fourier Transform (FFT) of the time series (see Figure \ref{fig:resultadoBiasTemporal2}), and of an artificial series normally distributed using the average and standard deviation of the real data. The frequency ranges from 0 to 0.0417 Hz ($\frac{1}{2} v_s$, being $v_s = \frac{1}{12}$, the sampling frequency), according to the Nyquist frequency (Blackledge 2006). Any frequency peak in the real data that is three sigma above the artificial data ($\bar M + 3\sigma$) is considered a detected signal, i.e., a periodic modulation in the bias. Figure \ref{fig:resultadoBiasTemporal2} presents an example of the FFT analysis, where the blue line shows the FFT of the time series and the red line shows the FFT of the artificial data. The goal is to detect periodic signals in the time series. For the same experiment performed on the four cameras, no periodic signal was found. 

{\renewcommand{\arraystretch}{1.2}
\renewcommand{\tabcolsep}{0.2 cm}
\begin{table*}[t]
\centering
\captionof{table}{Summary of the parameters obtained for the read noise characterization of the four SPARC4 CCDs.}
\label{tab:resultadoGradiente}
\begin{tabular}{cccccc}
\hline
\hline
Camera & $\bar{B}$\textsubscript{s}  & $\bar{B}$\textsubscript{t} & $\bar M$\textsubscript{r} & $\bar M$\textsubscript{r} $\times \; G$ & $\sigma$\textsubscript{man}\\
 & (ADU) & (ADU) & (ADU) & (e-) & (e-)\\
\hline
9914 & 498.2$\pm$0.4 & 498.4$\pm$0.4 & 1.86$\pm$0.47 & 6.32$\pm$1.60 & 6.66\\
9915 & 499.0$\pm$0.3 & 499.1$\pm$0.3 & 1.90$\pm$0.36 & 6.29$\pm$1.19 & 6.57\\
9916 & 500.0$\pm$0.2 & 500.0$\pm$0.3 & 1.91$\pm$0.16 & 6.30$\pm$0.53 & 6.67\\
9917 & 499.8$\pm$0.2 & 499.9$\pm$0.3 & 1.90$\pm$0.15 & 6.42$\pm$0.51 & 6.55\\
\hline
\end{tabular}
{\newline \footnotesize Note. $\bar{B}$\textsubscript{s} is the mean value obtained from the bias PDF presented in Figures \ref{fig:RN9914}-\ref{fig:RN9917}; $\bar{B}$\textsubscript{t} is the mean value of the time series presented in Figures \ref{fig:resultadoBiasTemporal1}; $\bar M$\textsubscript{r} is the mean value of the read noise obtained for the PDF presented in panel (b) of the Figures \ref{fig:RNspatial9914}-\ref{fig:RNspatial9917}. $G$ is the electronic gain presented in Table \ref{tab:resultadoGanho}, and $\sigma$\textsubscript{man} is the manufacturer's read noise in electrons.}
\end{table*}
}

\section{Electronic gain} \label{sec:GainCharact}
 The electronic gain of the CCD represents the conversion factor of the average number of photoelectrons necessary to generate one count. We apply the Janesick method (Janesick 2001)  to determine the gain. This method requires a series of flat and bias images, where the electronic gain is obtained through the relation between the signal intensity and its variance. The procedure adopted to obtain the gain is presented in Appendix A.

The flat images were obtained using a tungsten light source OL740-20A and a monochromator OL750-S Optronic Laboratories, INC. The goal is to produce a flat illumination on the CCD. The light source was turned on and set to 75~W, waiting 20 minutes for thermal stabilization. The camera was positioned on the monochromator output and the wavelength was adjusted to 1050 nm, for allowing a better signal intensity control. The CCD configuration used is:

\begin{myindentpar}{0.5cm}
1. Automatic shutter opening and closing \newline
2. Temperature of -70 ºC \newline
3. Exposure time: 3.5 s to 82 s \newline
4. Pre-amplification of 1\newline
5. Readout rate of 1 MHz\newline
6. Vertical shift speed of 4.33 x 10$^{-6}$ s \newline
7. Conventional Mode \newline
8. Read mode: Single Scan \newline
\end{myindentpar}

\begin{figure*}
\centering
\hspace{0pt}\includegraphics[scale=0.07]{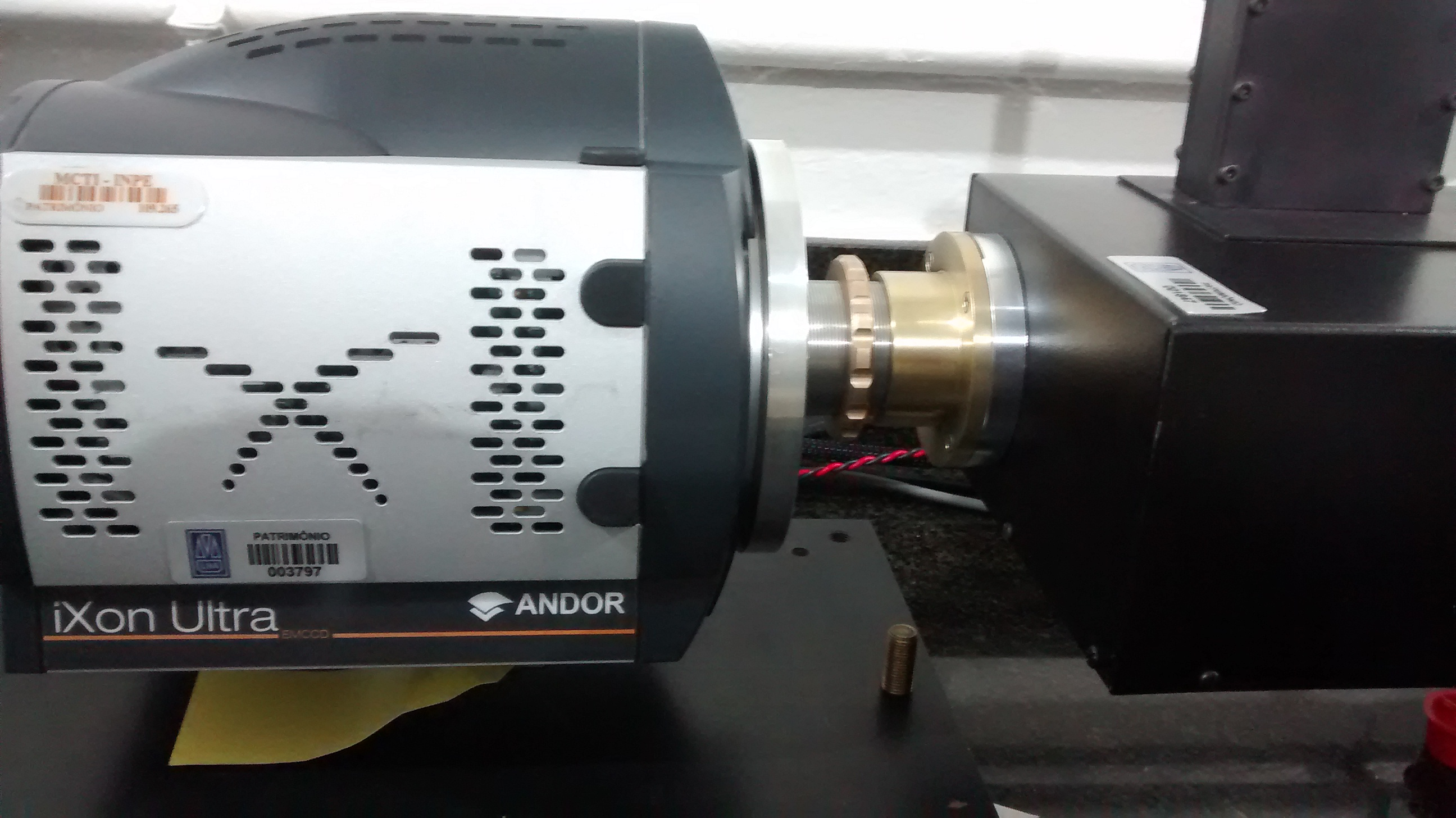}
\captionof{figure}{Camera mounted on the monochromator output.}
\label{fig:camera_montagem}
\end{figure*}

\begin{figure*}
\centering
\includegraphics[scale=0.4]{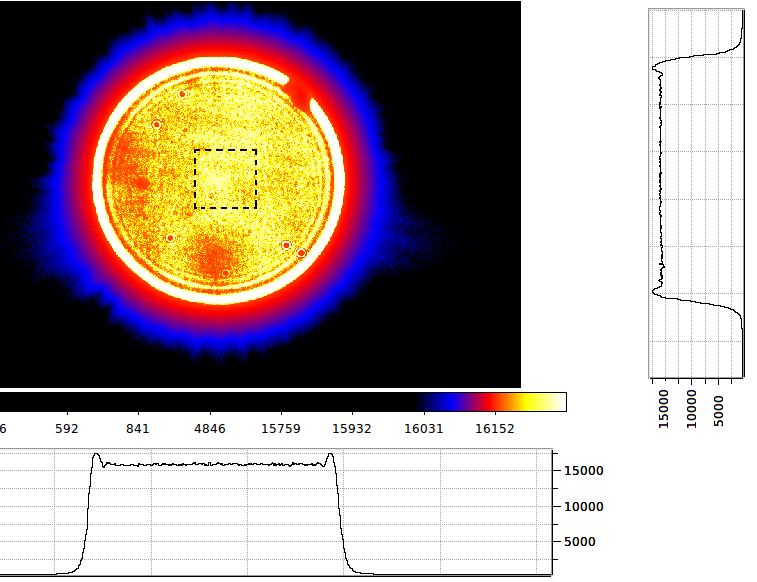}
\captionof{figure}{Incident light beam from the monochromator recorded by the CCD (output image of the DS9 Software version 7.2). The dashed line contour shows the region we considered for gain measurements. The bottom and right panels present the image flux profiles obtained at the center of the image, where one can see the flatness of the light spot.}
\label{fig:ds9}
\end{figure*}

Figure \ref{fig:camera_montagem} shows the CCD camera mounted on the monochromator. Figure \ref{fig:ds9} shows a sample flat image, where we also show the detector region considered for the gain measurements. After thermal stabilization, we take two bias images to clean the CCD. Then we obtain ten bias and five flats for ten different intensity levels. The intensity level is controlled by adjusting the exposure time keeping the light intensity constant. We kept the average of counts between 20~\% to 60~\% of the CCD dynamic range, ensuring that this value is well within the linear regime (Andor Technology 2015).

\begin{figure*}
\hspace{-30pt}\includegraphics[scale=0.27]{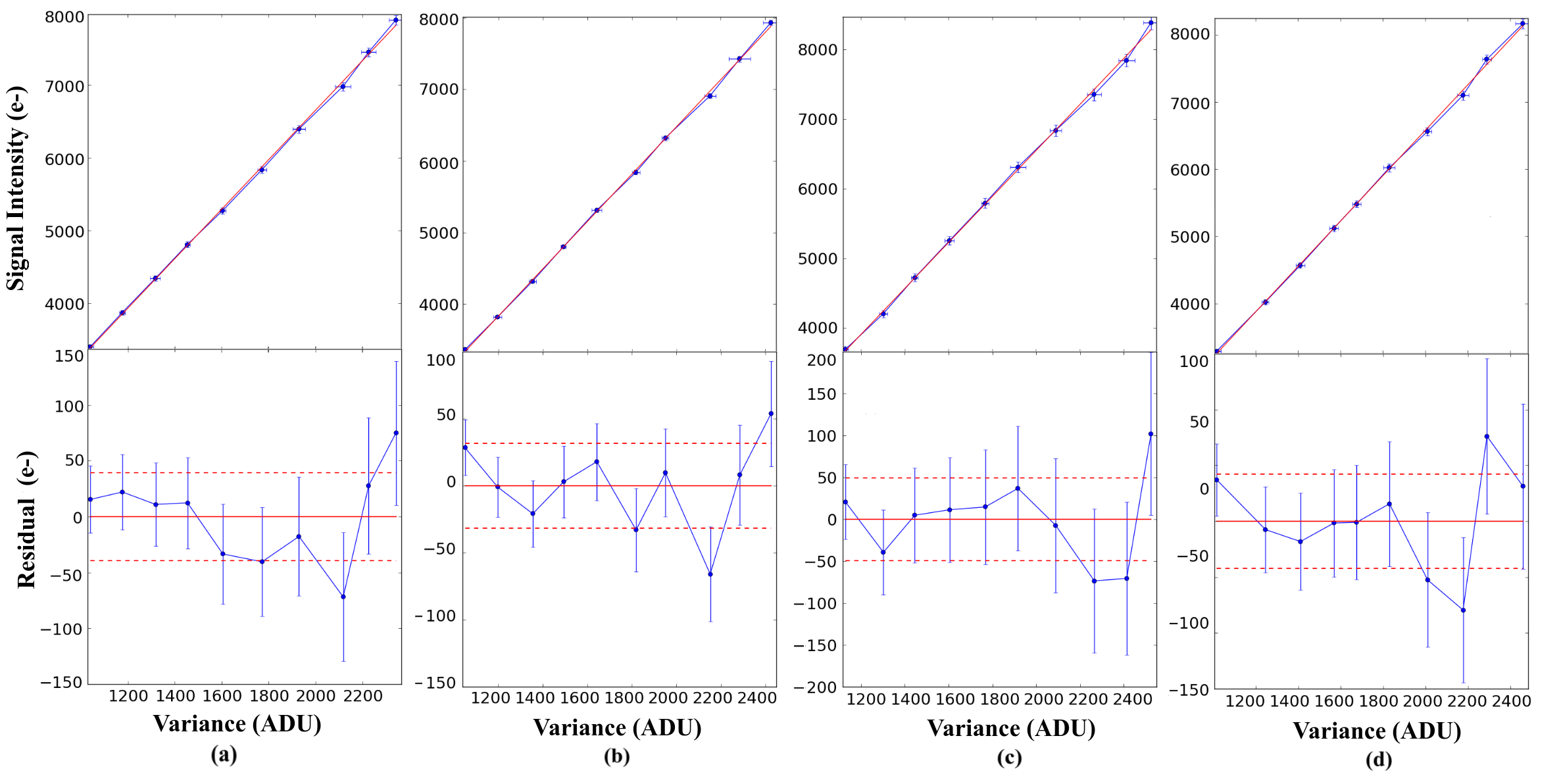}
\captionof{figure}{Top panels show the signal intensity vs. variance, i.e., the photon transfer curve (in blue) and the linear fit (in red). Bottom panels show the residuals. The panels show results for cameras 9914 (a), 9915 (b), 9916 (c), and 9917 (d).}
\label{fig:resultadoGanhoPlot}
\end{figure*}

Figure \ref{fig:resultadoGanhoPlot} presents the photon transfer curve and the best linear fit model for the four cameras. The measured values for the electronic gain and the manufacturer's values of the electronic gain are presented in Table \ref{tab:resultadoGanho}. The measured values of the gain are close to those provided by manufacturer.

{\renewcommand{\arraystretch}{1.2}
\renewcommand{\tabcolsep}{0.2 cm}
\begin{table*}
\centering
\captionof{table}{Summary of the electronic gain characterization: second column shows the electronic gain ($G$) values measured by the photon transfer curve.}
\label{tab:resultadoGanho}
\begin{tabular}{ccc}
\hline
\hline
Camera & $G$ & $G$\textsubscript{man}\\
 & (e-/ADU) & (e-/ADU)\\
\hline
9914 & $3.40 \pm 0.03$& 3.38\\
9915 & $3.31 \pm 0.03$& 3.30\\
9916 & $3.30 \pm 0.04$& 3.37\\
9917 & $3.38 \pm 0.03$& 3.36\\
\hline
\end{tabular}
{\newline \footnotesize Note. Third column shows the manufacturer's electronic gain ($G$\textsubscript{man}).}
\end{table*}}

\section{Dark current} \label{sec:DCcharact}
The dark current (DC) characterization is performed using a series of images taken with the shutter closed, using different exposure times and CCD temperatures. For temperatures of \newline -30 ºC and -40 ºC we acquired DC images with exposure times of 10 s, 5 minutes, 10 minutes, and 15 minutes. For temperatures of -50 ºC and -60 ºC the exposure times were 10 s, 10 minutes, 20 minutes, and 30 minutes. For temperature of -70 ºC, the exposure times were 10 s, 15 minutes, 30 minutes, 45 minutes, and 60 minutes. This procedure was adopted to increase the baseline of exposure times at low temperatures to improve the signal-to-noise ratio (SNR). In each series, we acquired ten bias images to subtract from the DC images. The CCD configuration used is:

\begin{myindentpar}{0.5cm}
1. Shutter constantly closed \newline
2. Temperature of -30, -40, -50, -60 and -70 ºC \newline
3. Pre-amplification of 1\newline
4. Readout rate of 1 MHz\newline
5. Vertical shift speed of 4.33 x 10$^{-6}$ s \newline
6. Conventional Mode \newline
7. Read mode: Single Scan \newline
\end{myindentpar}

The data analysis is divided into the following three steps: DC determination, the dependency of the DC with internal CCD temperature, and the spatial distribution of the DC. 

\begin{figure*}
\centering
\includegraphics[scale=0.3]{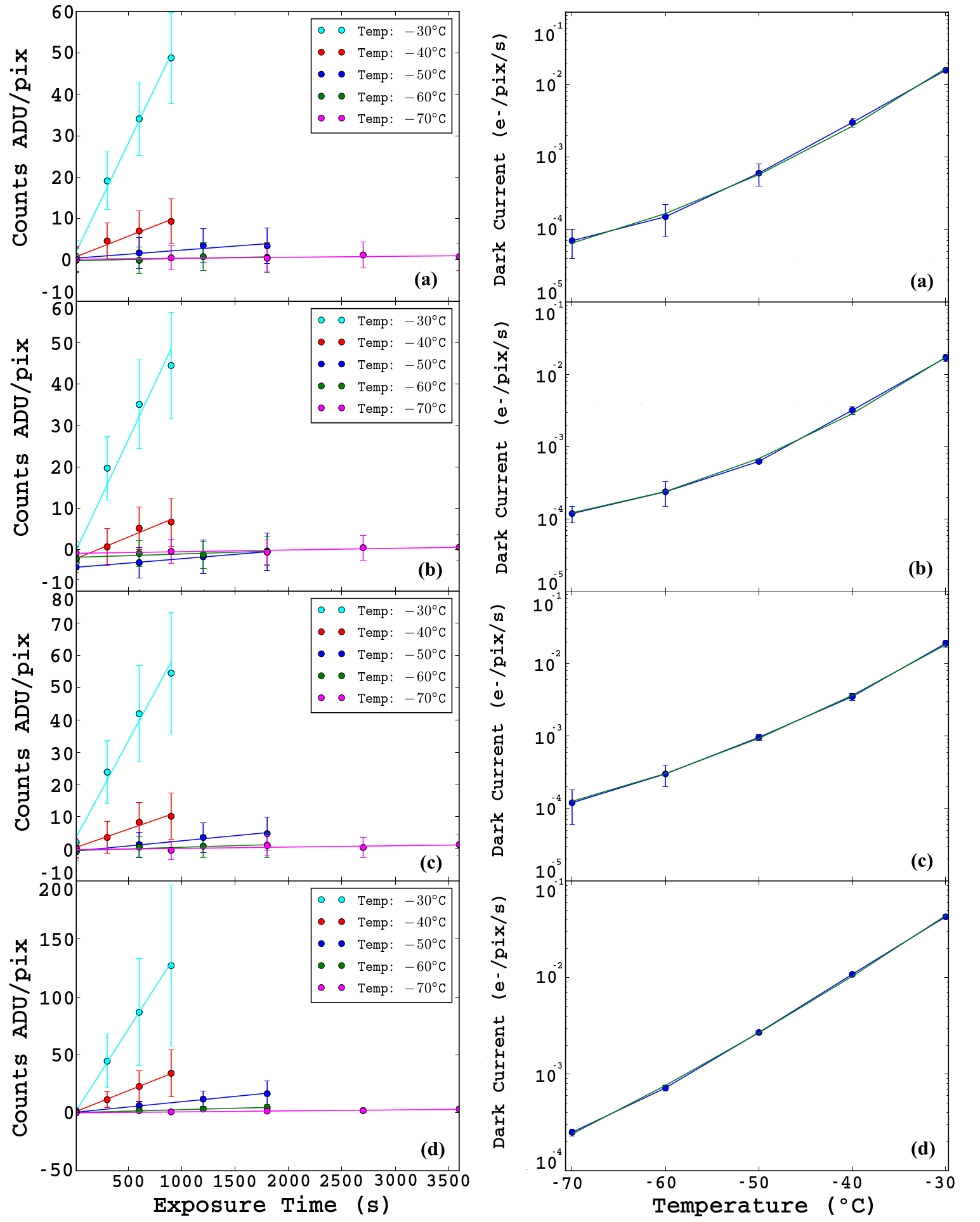}
\captionof{figure}{Dark current characterization for cameras 9914 (a), 9915 (b), 9916 (c) and 9917 (d): left-hand side panels show the median of counts (ADU) as a function of exposure time; right-hand side panels show the dark current values (blue line) in a log scale as a function of temperature and the fit model (green line) as described in the text.}
\label{fig:resultadoDCtemporal}
\end{figure*}

The DC is given by the slope of a linear fit to the median number of electrons/pix ($n_{e}$) generated in a given dark exposure as a function of exposure time ($t$\textsubscript{exp}), where $n_e$ is calculated as follows:

\begin{equation}
n_{e} = (D - B) \times G
\end{equation} 

\noindent for $D$ being the median of counts (in ADU) in a dark exposure, $B$ being the median bias level in ADU, and $G$ being the electronic gain in \newline e-/ADU. This process is repeated for each temperature, allowing us to observe the DC behavior as a function of the CCD temperature. Figure~\ref{fig:resultadoDCtemporal} presents the measured DC for the four SPARC4 cameras. Left-hand side panels show the median of counts as a function of exposure time and the linear fit to obtain the DC. Right-hand side panels show the DC in log scale, as a function of CCD temperature. 

Table \ref{tab:resultadoDCTemp} presents the measured values of DC. Table \ref{tab:manufacturerDC} presents the DC values provided by the manufacturer. Our measurements confirm the low dark current in our detectors.  We fit the following parabolic model to the DC values as a function of temperature:

\begin{equation}
\hspace{-20pt} \ln{(DC)} = aT^2 + bT + c,
\label{eq:dcmodel}
\end{equation}

\noindent where $T$ is the temperature in Celsius and $a, b$ and $c$ are the fit coefficients. Equation \ref{eq:dcmodel} is an empiric model, thus it is only valid within the temperature range where it has been measured, i.e., between -30~ºC and -70~ºC. Therefore, we obtained the following DC models, in exponential form, for the four SPARC4 cameras:

\begin{equation}
\hspace{-20pt} DC_{\rm 9914} = 24.66 \; e^{0.0015 T^2 + 0.29 T}
\end{equation}

\begin{equation}
\hspace{-20pt} DC_{\rm 9915} = 35.26 \; e^{0.0019 T^2 + 0.31 T}
\end{equation}

\begin{equation}
\hspace{-20pt} DC_{\rm 9916} = 9.67 \; e^{0.0012 T^2 + 0.25 T}
\end{equation}

\begin{equation}
\hspace{-20pt} DC_{\rm 9917} = 5.92 \; e^{0.0005 T^2 + 0.18 T}
\end{equation}

{\renewcommand{\arraystretch}{1.2}
\renewcommand{\tabcolsep}{0.2 cm}
\begin{table*}[t]
\centering
\caption{Dark current values obtained by the slope of curves presented in Figure \ref{fig:resultadoDCtemporal}, multiplied by the gain values presented in Table \ref{tab:resultadoGanho}.}
\label{tab:resultadoDCTemp}
\begin{tabular}{ccccc}
\hline
\hline
Temperature (ºC) & \multicolumn{4}{c}{Dark Current $\times10^{-4}$ e-pix\textsuperscript{-1}s\textsuperscript{-1}} \\
\cline{2-5}
 & 9914 & 9915 & 9916 & 9917 \\
\hline
-30 & 158 $\pm$ 7 & 170 $\pm$ 20 & 190 $\pm$ 20 & 426 $\pm$ 7 \\
-40 & 30 $\pm$ 4 & 32 $\pm$ 4 & 35 $\pm$ 4 & 108 $\pm$ 2 \\
-50 & 6 $\pm$ 2 & 6.3 $\pm$ 0.2 & 9.6 $\pm$ 0.9 & 27 $\pm$ 1\\
-60 & 1.5 $\pm$ 0.7 & 2.4 $\pm$ 0.9 & 3 $\pm$ 1 & 7.2 $\pm$ 0.5 \\
-70 & 0.7 $\pm$ 0.3 & 1.2 $\pm$ 0.3 & 1.2 $\pm$ 0.6 & 2.5 $\pm$ 0.2 \\
\hline
\end{tabular}
\end{table*}}

{\renewcommand{\arraystretch}{1.2}
\renewcommand{\tabcolsep}{0.2 cm}
\begin{table*}[t]
\centering
\caption{Manufacturer dark current values.}
\label{tab:manufacturerDC}
\begin{tabular}{ccc}
\hline
\hline
CCD & Temperature & Manufacturer's DC\\
 & (ºC) & $\times10^{-4}$ e-pix\textsuperscript{-1}s\textsuperscript{-1}\\
\hline
9914 & -100.70 &  1.2 \\
9915 & -100.94 &  0.76\\
9916 & -101.41 & 1.1 \\
9917 & -101.26 & 1.2 \\
\hline
\end{tabular}
\end{table*}}

\begin{figure*}
\centering
\hspace{-60pt}\includegraphics[scale=0.3]{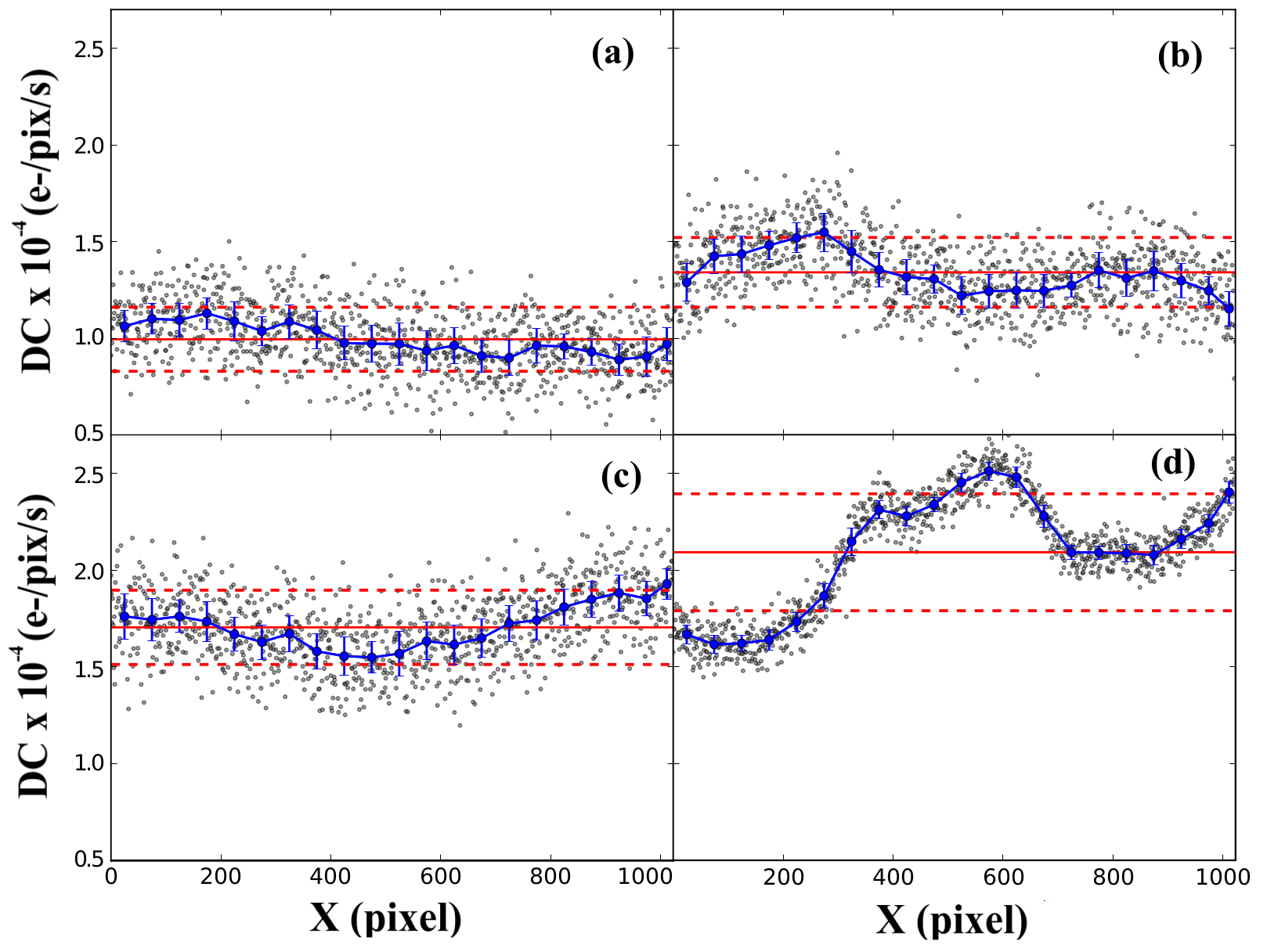}
\captionof{figure}{Spatial distribution of the dark current along the pixel columns for the CCDs 9914 (a), 9915 (b), 9916 (c) and 9917 (d). Gray dots present the calculated values of dark current for each CCD column. Blue lines show the binned dark current values along columns. Red lines show the mean (solid line) and one-sigma values (dashed lines).}
\label{fig:DCeixoX}
\end{figure*}

\begin{figure*}
\centering
\hspace{-60pt}\includegraphics[scale=0.3]{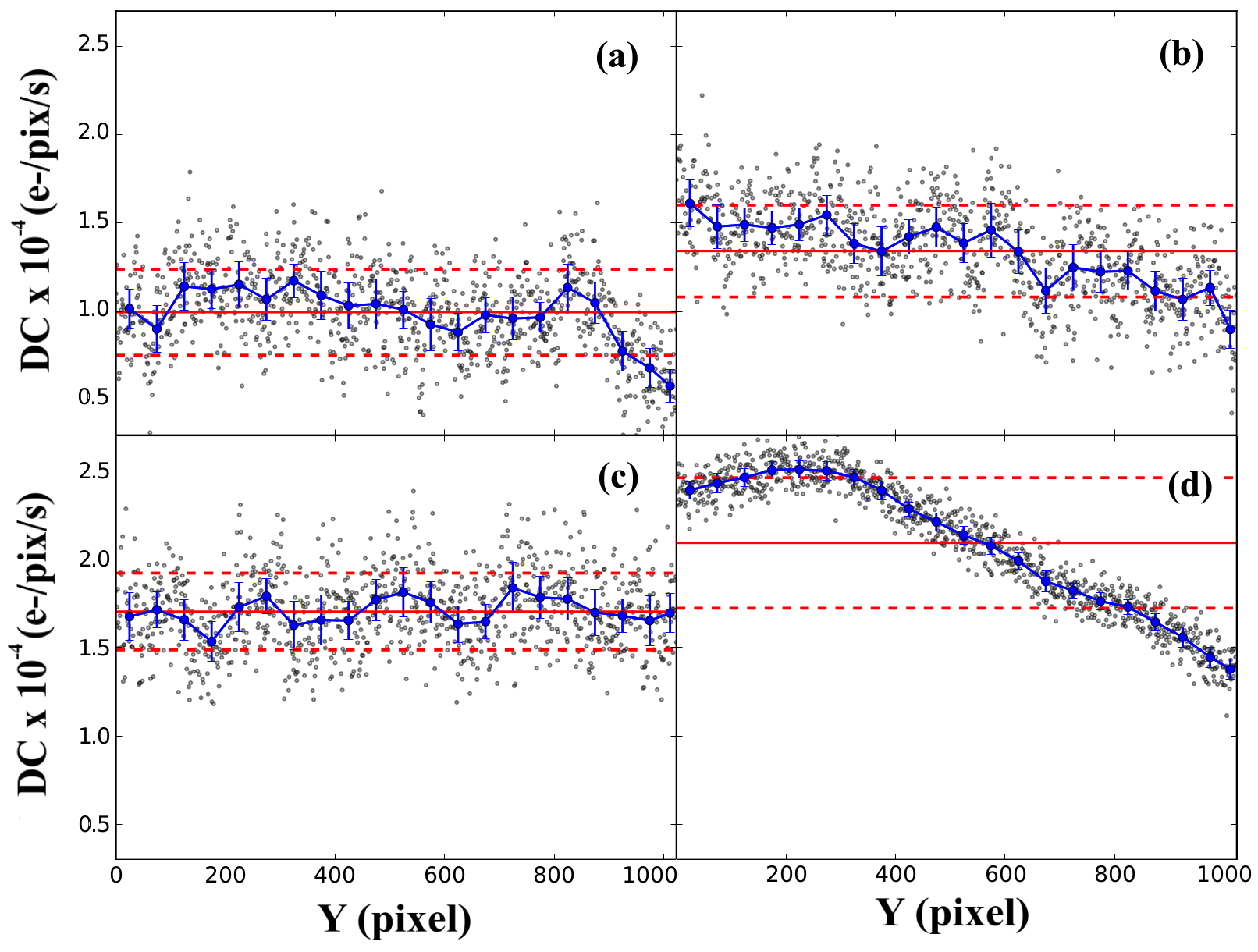}
\captionof{figure}{Spatial distribution of the dark current along the pixel rows for the CCDs 9914 (a), 9915 (b), 9916 (c) and 9917 (d). Gray dots present the calculated values of dark current for each CCD row. Blue lines show the binned dark current values along rows. Red lines show the mean (solid line) and the one-sigma values (dashed lines).}
\label{fig:DCeixoY}
\end{figure*}

The spatial distribution of DC was obtained only at temperature of -70 ºC. First we calculate the average of counts for each CCD column as a function of exposure time. Then we obtain the DC for each column by the linear fit method as described previously. This procedure is repeated for every CCD column, allowing us to verify the DC behavior over the {\it x}-direction (see Figure \ref{fig:DCeixoX}). We calculate the mean and standard deviation of the DC for every bin of 50 points, allowing us to evaluate if there is any systematic trend in the data. A similar procedure is performed for the CCD rows, allowing us to evaluate the DC behavior over the {\it y}-direction (see Figure \ref{fig:DCeixoY}). Based on Figure \ref{fig:DCeixoX} and Figure \ref{fig:DCeixoY}, one may notice that the camera 9917 presents the most DC spatial variation and also the highest mean DC value.

\section{Quantum efficiency (QE)} \label{sec:QECharact}

QE is defined as:

\begin{equation}
QE = \frac{N_e}{N_p}.
\end{equation}

\noindent It relates the number of photoelectrons ($N_e$) produced for a given number of incident photons ($N_p$) in a CCD pixel.

In our experiment, we used a photometer detector the measure the light flux emitted by a monochomator in the spectral range between 350 nm to 1100 nm. This result is used to compare the signal acquired by the CCD camera over the same spectral range, and then to determine the quantum efficiency of the CCD. 

\begin{figure*}
\centering
\hspace{0pt}\includegraphics[scale=0.08]{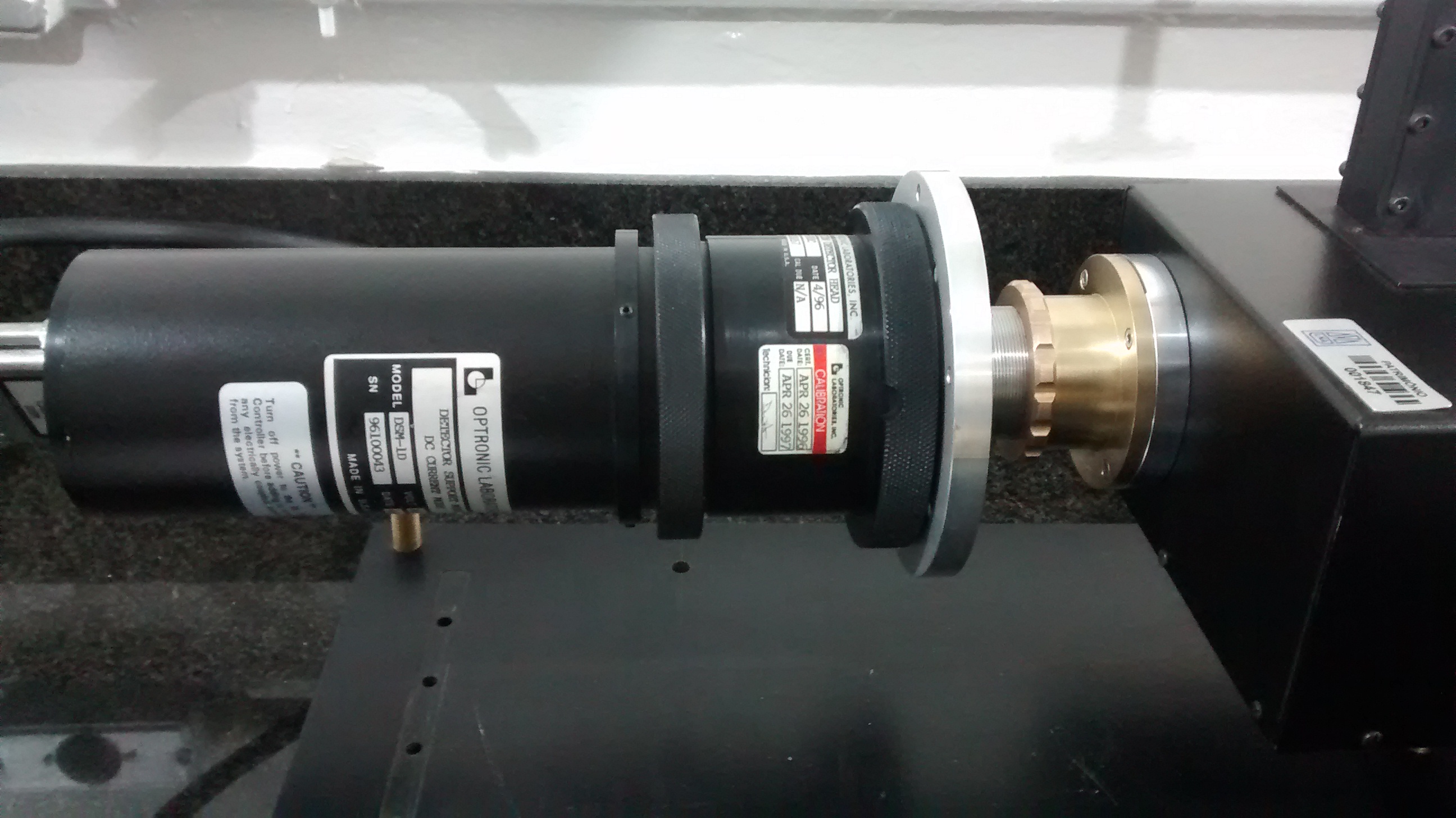}
\captionof{figure}{Picture of the photometer device mounted on the monochromator output.}
\label{fig:montagemFotometor2}
\end{figure*}

\begin{figure*}
\centering
\hspace{0pt}\includegraphics[scale=0.4]{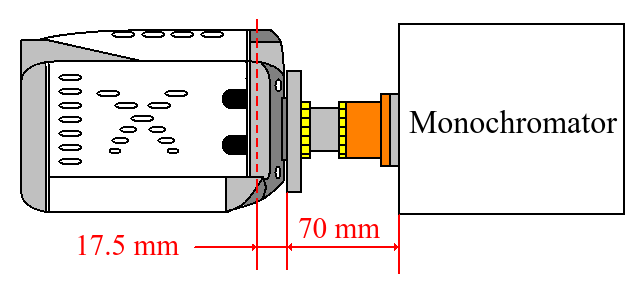}
\captionof{figure}{Distances of the CCD mount on the monochromator output.}
\label{fig:planoCCD}
\end{figure*}

\begin{figure*}
\centering
\hspace{0pt}\includegraphics[scale=0.4]{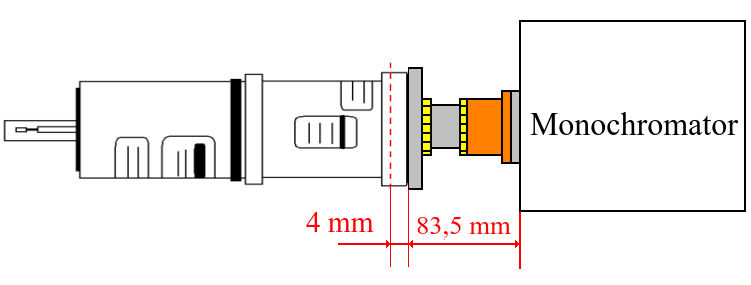}
\captionof{figure}{Distances of the photometer mount on the monochromator output.}
\label{fig:planofotometro}
\end{figure*}

The light source was turned on and set to 75 W, waiting 20 minutes for thermal stabilization. Then, the photometer was positioned on the monochromator output using a custom flange adaptor (see Figure \ref{fig:montagemFotometor2}), which allows us to place the photometer at the same distance as the CCD, ensuring the same light flux. The CCD has also a custom flange adaptor (see Figure \ref{fig:camera_montagem}).  Figures \ref{fig:planoCCD} and \ref{fig:planofotometro} illustrate how the CCD and the photometer are mounted on the monochromator output, respectively. We performed measurements in the spectral range between 350 nm and 1100 nm, with steps of 50 nm. Immediately after the photometer measurements, we remove it and mount the CCD, where we perform measurements in the same spectral range. For the spectral region around 650 nm, the CCD saturates even using the shortest possible exposure time. Therefore we used a density filter ({\it D}=2.0), where the transmission curve of the filter is taken into account in our analysis (more details in Appendix B).

The equipment used in the experiment was a photometer OL750-HSD-301C high-sensitivity, a tungsten lamp OL750-20A, and a monochromator OL750-S.The CCD configuration used is presented below:

\begin{myindentpar}{0.5cm}
1. Automatic shutter opening and closing \newline
2. Temperature of -70 ºC \newline
3. Pre-amplification of 1\newline
4. Readout rate of 1 MHz\newline
5. Vertical shift speed of 4.33 x 10$^{-6}$ s \newline
6. Conventional Mode \newline
7. Read mode: Single Scan \newline
\end{myindentpar}

We obtained the incident power in Watts with the photometer, which is divided by the photon energy, $E_{\lambda}=hc/\lambda$, to obtain the number of photons per second (according to the IS), i.e., 

\begin{equation}
N_{p} = \frac{P}{E_{\lambda}} = \frac{I\lambda}{hcR_{\lambda}}.
\end{equation}

The number of electrons generated per second in the CCD is obtained by the expression:

\begin{equation}
N_{e} = \frac{S*G}{t_{\rm exp}}
\end{equation}

\noindent where $S$ is the sum of counts measured in the image in units of ADU, $G$ is the CCD gain in \newline e-/ADU, and $t_{\rm exp}$ is the exposure time in seconds.  The above calculations assume that the area of the photometer is the same as the area of the CCD. For this reason, we have measured the size of the photometer projected on the CCD array, and then we only considered the image pixels within the photometer footprint on the CCD. 

\begin{figure*}[t]
\centering
\includegraphics[scale=0.4]{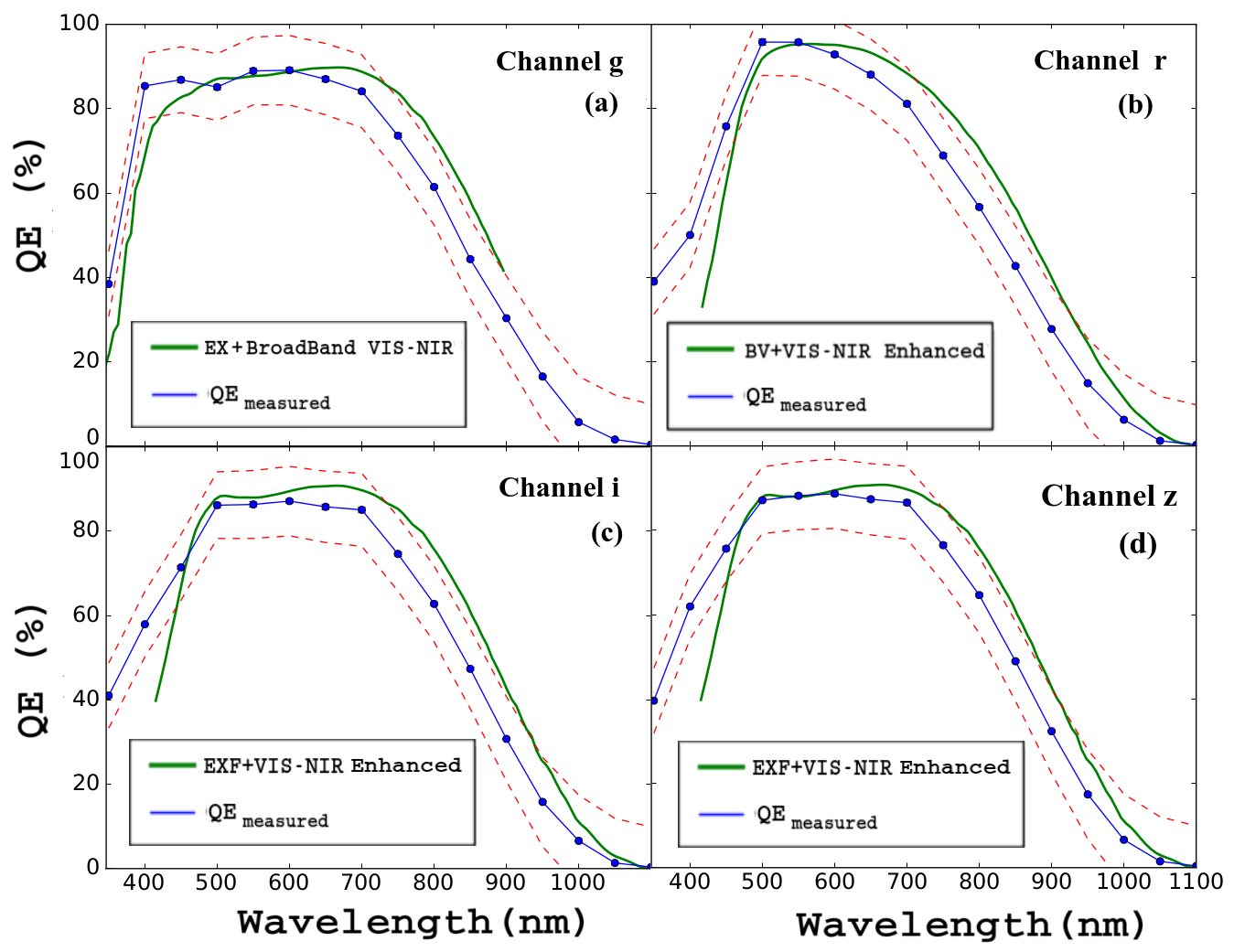}
\captionof{figure}{Quantum efficiency for CCDs 9914 (a), 9915 (b), 9916 (c) and 9917 (d) at -70 ºC. Blue line shows the measured quantum efficiency and red dashed lines show the estimated uncertainty from mechanical misalignment and light source stabilization. Green line presents the manufacturer's quantum efficiency. The legend presents the detector and coating specifications.}
\label{fig:resultadoEQ}
\end{figure*}

Figure \ref{fig:resultadoEQ} presents the results for the QE measurements for all the four SPARC4 cameras. The QE in our experiment was obtained with CCD temperature of -70 ºC. Our measurements (blue lines) and uncertainties (red dashed lines) are compared with the manufacturer's QE (green lines), which were obtained with CCD temperature of 25 ºC (Andor Technology 2017). The manufacturer values take into account the specific window and coating of each detector. The legend in Figure \ref{fig:resultadoEQ} shows these specifications. Notice that QE increases with CCD temperature (Lesser and McCarthy 1996), especially for redder wavelengths. This fact explains the differences between our measured QE and the manufacturer's QE, similar to the results presented by O'Connor et al. (2008). Table \ref{tab:resultadoQE} shows the maximum QE measured from our data, the wavelength of the maximum QE, and the total percentage of photon flux converted into electrons for the range between 350 nm and 1100 nm. \ref{sec:QEvalues} presents the values of the QE along the spectrum for the four cameras.

{\renewcommand{\arraystretch}{1.2}
\renewcommand{\tabcolsep}{0.1 cm}
\begin{center}
\captionof{table}{Parameters of the quantum efficiency characterization as presented in Figure \ref{fig:resultadoEQ}: maximum quantum efficiency (QE\textsubscript{max}), wavelength for the maximum quantum efficiency ($\lambda$\textsubscript{max}) and the total quantum efficiency ($T$\textsubscript{QE}) percentage of photon flux absorbed over the spectral range between 350 nm and 1100 nm.}
\label{tab:resultadoQE}
\begin{tabular}{cccc}
\hline
\hline
Camera & QE\textsubscript{max} & $\lambda$\textsubscript{max} & $T$\textsubscript{QE}\\
& (\%) & (nm) & (\%) \\
\hline
9914 & 89.1 & 600 & 57.79\\
9915 & 95.8 & 500 & 54.49\\
9916 & 87.1 & 600 & 54.78\\
9917 & 88.7 & 600 & 56.40\\
\hline
\end{tabular}
\end{center}}

\section{Conclusions} \label{sec:conclusao}

We presented the experimental methodologies to characterize the read noise, electronic gain, dark current and quantum efficiency for the four SPARC4 Ixon Ultra 888 CCDs developed by Andor Technology.  In our experiments, we verified that the spatial distribution of the bias level for cameras 9914 and 9915 are not uniform.  However, the spatial distribution of the read noise appears to be homogeneous for all cameras. Moreover, we verified that the temporal and spatial variability of the bias are $\sim 8$ times lower than the electronic read noise. We detected a bias drift for all CCDs of approximately 0.6 ADU in two hours for an acquisition frequency of $\sim0.1$\,Hz. Notice that this drift should also be present in science acquisition for continuous time series. The cameras 9914, 9915, and 9916 did not present a spatial variability of the DC. However, camera 9917 presented a higher level of DC gradient than in the other cameras. We have also inspected the variation of the DC as a function of CCD temperature, which shows that for temperatures below $\sim -50^o$C, the variation of the DC is small. We have also measured the electronic gain of the CCD with uncertainties of 1\%, which is important for the conversion of measured flux into physical units. Finally, we have determined the QE for the spectral range between 350 nm and 1100 nm with uncertainties of $\lesssim$ 15 \%. These results show the actual efficiency for each of the four SPARC4 cameras.

Overall, the results presented here are comparable with those provided by the manufacturer, which makes us believe that our methodology is reliable. Besides the application to the other modes of operations of the SPARC4 CCDs, our methods can also be extended to the characterization of other CCDs available at OPD, which do not have enough information from the manufacturer. 

Our experiments provide a way to measure the CCD performance parameters systematically. Although we only presented results for one mode of operation, it is now possible to quickly repeat the same experiments to characterize the other modes of operation, where one may select different CCD readout rates, pre-amplifier gain, and also the electron multiplying mode, which may have different noise characteristics. Also it is useful to monitor the characteristics of our CCDs over time to inspect how the aging of our cameras is affecting their performances. Python language, version 2.7.6, was used to develop the codes to perform data reduction. A complete tutorial with explanations on how to perform the measurements presented in this paper are available at \url{wiki.lna.br/wiki/opdccds}.  

\ack
We want to thank Rodrigo Prates Campos, Saulo Roberly Gargaglioni and Adriano Messala Coimbra for their technical support with the equipments. We want to thank the LNA's mechanics workshop staff for machining the parts necessary to the electronic gain and QE experiments. CVR would like to thank Fapesp for financial support under the thematic Project 13/26258-4 and CNPq (Proj:306701/2015-4). DVB want to thank MCTIC for the internship funds under program nº 01204.000110/2016-13, and CNPq for the PCI funds under program nº 304660/2017-5.

\appendix

\section{The Janesick method} \label{sec:Janesick}
The Janesick method is based on the relation between the average signal ($S$) acquired by the CCD and its variance ($\sigma_S^2$) (Janesick 2001). Let us consider the following equation: 

\begin{equation}
\sigma_S^2 =  \frac{\sigma_p^2}{G} + \sigma_R^2  \label{eq:gainEQ},
\end{equation}

\noindent where the variance of the signal $\sigma_S^2$ is given by the sum of the variance of the photometric error, $\sigma_p^2$, divided by the gain $G$, plus the variance of the read noise $\sigma_R^2$ in ADU, disregarding the correlation terms.

It is assumed that the photometric error is composed only by the photon noise, which follows the Poisson distribution, where $\sigma_p^2=S$ (Janesick et al. 1993). Thus, the CCD gain is given by:

\begin{equation}
G = \frac{S}{\sigma_S^2 - \sigma_R^2} \label{eq:gainEQ2}.
\end{equation} 

The read noise in electrons ($\sigma_e$) is related to the read noise in ADU ($\sigma_{R}$) by the following equation: 

\begin{equation}
\sigma_e = G * \sigma_{R} \label{eq:sigmaE}.
\end{equation}

The photon transfer curve is a graphic method to obtain the gain value $G$. It relates the average intensity signal as a function of its variance. The procedure is given by obtaining pairs of bias images ($B_{A}, B_{B}$) and pairs of flat images ($F_{A}, F_{B}$) acquired for different intensity levels. Where the average signal for each pair of flats is given by:

\begin{equation}
\hspace{-50pt}S = \big<F_A\big> + \big<F_B\big> - \big<B_A\big> - \big<B_B\big>,
\label{eq:averagesignal}
\end{equation}

\noindent for $\big<F_A\big>$ and $\big<F_B\big>$ being the average of each flat-field image in the pair obtained for the same intensity level, and $\big<B_A\big>$ and $\big<B_B\big>$ being the average of two bias images. The read noise for the pair of bias, $\sigma_{\Delta B}$, is given by:

\begin{equation}
\sigma_{\Delta B} = \sqrt{2}\sigma_{R} \label{eq:readnoisepairbias}.
\end{equation}

Through these pair of images we find a linear relation, from which the gain and read noise values can be obtained (Christen 2007).  Combining equations \ref{eq:gainEQ2}, \ref{eq:sigmaE}, and also considering equations \ref{eq:averagesignal} and \ref{eq:readnoisepairbias}, one finds:

\begin{equation}
Y = X  G  - \sigma_e,
\end{equation}

\noindent where {\it Y} is given by:

\begin{equation}
\hspace{-50pt}Y = \frac{\big<F_A\big> + \big<F_B\big> - \big<B_A\big> - \big<B_B\big>}{\sqrt{2} \; \sigma_{\Delta B}},
\end{equation}

\noindent and the {\it X} value is given by:

\begin{equation}
X = \frac{\sigma_{\Delta F}^2}{\sqrt{2} \; \sigma_{\Delta B}},
\end{equation}

\noindent where we considered $\sigma_{S}^2 = \sigma_{\Delta F}^2$, which is the variance of the difference between the two flat images. This procedure is repeated for different intensity levels. We perform a linear fit to these data, where the gain is given by slope and the read noise is given by the intercept.

\section{Transmission curve of the density filter} \label{ssec:densityFilter}
During the QE experiments, the incident light saturates the CCD even for the shortest possible exposure times. So, we used a density filter to limit the intensity of the incident light. In data processing, it is necessary to consider the transmission curve of the filter. Table \ref{tab:densityFilter} presents the transmission values of the filter used in the QE experiment. 

{\renewcommand{\arraystretch}{1.0}
\renewcommand{\tabcolsep}{0.2 cm}
\begin{center}
\captionof{table}{Transmission values of the filter in the spectral interval 400nm to 1050 nm used in the quantum efficiency experiment; Tr represents the percentage of the transmission of the filter according to the wavelength.}
\label{tab:densityFilter}
\begin{tabular}{cc}
\hline
\hline
$\lambda $ & Tr\\
 (nm) & (\%) \\
\hline
400 & 0.53\\
450 & 1.09\\
500 & 1.10\\
550 & 1.15\\
600 & 1.15\\
650	& 1.60\\
700	& 3.33\\
750	& 4.90\\
800	& 5.49\\
850	& 5.44\\
900	& 5.06\\
950	& 4.68\\
1000 & 4.50\\
1050 & 4.54\\
\hline	
\end{tabular}
\end{center}}

\section{Quantum efficiency values for the SPARC4 cameras} \label{sec:QEvalues}
This section presents the values (Table \ref{tab:QEValues}) obtained by the QE characterization of the SPARC4 cameras, presented in Figure \ref{fig:resultadoEQ} . 

{\renewcommand{\arraystretch}{1.0}
\renewcommand{\tabcolsep}{0.2 cm}
\begin{center}
\captionof{table}{Quantum efficiency values obtained by the quantum efficiency experiment described in Section  \ref{sec:QECharact}.}
\label{tab:QEValues}
\begin{tabular}{ccccc}
\hline
\hline
$\lambda$ (nm) & \multicolumn{4}{c}{QE (\%)}\\
\cline{2-5}\\
 & 9914 & 9915 & 9916 & 9917\\
\hline
350.0 &	38.43 & 38.98 & 41.01 & 39.60\\
400.0 &	85.39 & 50.04 & 57.93 & 61.93\\
450.0 &	86.86 & 75.82 & 71.43 & 75.66\\
500.0 &	85.12 & 95.80 & 86.17 & 87.12\\
550.0 &	88.96 & 95.77 & 86.34 & 88.20\\
600.0 &	89.12 & 92.86 & 87.13 & 88.67\\
650.0 &	87.01 & 88.07 & 85.78 & 87.36\\
700.0 &	84.12 & 81.16 & 85.07 & 86.55\\
750.0 &	73.60 & 68.90 & 74.61 & 76.48\\
800.0 &	61.44 & 56.67 & 62.76 & 64.63\\
850.0 &	44.32 & 42.70 & 47.36 & 48.95\\
900.0 &	30.32 & 27.75 & 30.75 & 32.34\\
950.0 &	16.49 & 14.90 & 15.80 & 17.37\\
1000.0 & 5.67 &  6.24 &  6.58 &  6.60\\
1050.0 & 1.56 &  1.25 &  1.33 &  1.49\\
1100.0 & 0.35 &  0.27 &  0.35 &  0.38\\	
\hline
\end{tabular}
\end{center}}

\end{multicols}

\section*{References}
\begin{thebibliography}{99}
\bibitem{1} Akhlaghi M., 2017. GNU Astronomy Utilities, version 0.5. Free Software Foundation (FSF), https://www.gnu.org/software/gnuastro/

\bibitem{2} Andor Technology, 2017. iXon Ultra - The World’s Highest Performance Back-illuminated EMCCDs. Belfast, BT12 7AL, United Kingdom, http://www.andor.com/

\bibitem{3} Andor Technology, 2015. iXon Ultra 888 - Hardware Guide, ed. 1.0, Belfast BT12 7AL, United Kingdom, http://www.andor.com/

\bibitem{4} Blackledge, J. M., 2006. Digital Signal Processing: Mathematical and Computational Methods, Software Development and Applications. Ed. 2,Horwood Publishing Limited, Chichester PO20 3QL, England, 840

\bibitem{5} Christen, F.~F.~T. 2007, Ph.D. thesis, Univ. Groningen

\bibitem{7} Gunn, J.~E.; Carr, M.; Rockosi, C., et al. 1998, AJ, 116, 3040

\bibitem{9} Janesick, J.~R.\ 2001, Scientific charge-coupled devices (Bellingham, WA: SPIE Optical Engineering Press), 906

\bibitem{8} Janesick, J.~R., Elliott, T., Collins, S., Blouke, M.~M., \& Freeman, J.\ 1993, Selected Papers on Instrumentation in Astronomy (Bellingham, WA: SPIE Optical Engineering Press), 513.

\bibitem{10} Lesser, M.~P., \& McCarthy, B. 1996, Proc. SPIE 2654, 278

\bibitem{11} O'Connor, P., Frank, J., Geary, J.~C., et al. 2008, Proc. SPIE, 7021, 702106

\bibitem{12} Rodrigues, C.~V.; Taylor, K.; Jablonski, F.~J., et al.\ 2012, Proc. SPIE, 8446, 844626
\end{thebibliography}
\end{document}